\documentclass[]{jaa}
\usepackage{natbib}
\usepackage{graphicx}
\usepackage{multirow}
\usepackage{multicol}
\usepackage{hhline}
\usepackage{xcolor}
\usepackage{varioref}
\usepackage{stackengine}
\usepackage{csquotes}
\usepackage{ifthen}
\usepackage{url}
\usepackage{euscript}
\usepackage{nameref}

\begin{document}\sloppy

\title{Statistical Investigation of B-fields in Cores and Filaments Using JCMT/SCUPOL Legacy Survey Archival Data}


\author{Puja Porel\textsuperscript{*}}
\affilOne{Indian Institute of Astrophysics, Bengaluru, 560034, India\\}


\twocolumn[{

\maketitle

\corres{puja.porel@iiap.res.in}


\begin{abstract}
SCUPOL, the polarimeter for SCUBA on the James Clerk Maxwell Telescope, was used for polarization observations of 104 regions at 850 $\mu$m wavelength and 15" resolution in the mapping mode by \cite{2009ApJS..182..143M}. They presented the polarization values and magnetic field morphologies in these regions. In this work, we took the opportunity to use this big legacy survey data to investigate further the collective statistical properties of the measured polarization in different star-forming regions containing cores and filaments. We did not reproduce the polarization maps but used the polarization value catalogs to investigate the statistics of distributions. In some of these regions, the data from Combined Array for Research in Millimeter-wave Astronomy (CARMA) polarization observation at 1.3 mm wavelength and 2."5 resolution was also available from the TADPOL survey \citep{2014ApJS..213...13H}. We used that data and compared it with JCMT/SCUPOL values. We also study how the direction of outflows appears to relate the mean B-field direction from large scale (JCMT observation at 15") to small scale (CARMA observation 2.5") for the nine core regions common in both.

\end{abstract}
\keywords{ISM---Dust: Polarization---Magnetic fields.}

}]



\section{Introduction}

Magnetic fields are crucial components in the star formation process \citep{liu2022magnetic:, li2014link}, significantly influencing the dynamics involved \citep{1987ARA&A..25...23S, 1993prpl.conf..327M}. The study of magnetic field morphology within the interstellar medium has been extensively advanced through observations of polarization caused by the alignment of interstellar dust grains. This phenomenon was initially discovered by \cite{1949Sci...109..166H}, and \cite{1949ApJ...109..471H}, \cite{1949Sci...109..165H}. For many years, the prevailing theory was that paramagnetic relaxation was responsible for aligning these rapidly rotating dust grains with the magnetic field. However, this alignment mechanism has faced significant challenges, both observationally \citep{2008MNRAS.387..797H} and theoretically \citep{2007MNRAS.378..910L}, revealing numerous limitations in its explanatory power.

The ’Radiative Alignment Torque’ (RAT) theory is
currently the most widely accepted mechanism for the
alignment of interstellar dust grains. Initially proposed by \cite{1976Ap&SS..43..291D} and later fully developed by \cite{1996ApJ...470..551D} and \cite{2007MNRAS.378..910L}, this theory posits that in the presence of anisotropic radiation, the transfer of torques from photons to paramagnetic and non-spherical or elongated dust grains induces a rapid spin-up of these grains. As a result of this rotation, the grains acquire a net magnetic moment through the Barnett effect, causing them to precess around the magnetic field. This
magnetic moment further drives the angular momentum
of the grains to precess around the magnetic field
in a manner known as Larmor precession, ultimately
aligning the grains’ angular momentum parallel to the
magnetic field lines. Over the past decade, numerous predictions of this theory have been substantiated \citep{2015ARA&A..53..501A}, though the finer details of the mechanism and its components remain areas for further investigation.

When dust grains aligned by the interstellar magnetic field absorb radiation at shorter wavelengths, they re-emit this energy at longer wavelengths, specifically in the far infrared, millimeter, or sub-millimeter regions of the spectrum. The emitted radiation is polarized
along the grains’ long axis, meaning the electric field
vector of the polarized emission is perpendicular to the plane-of-sky component of the local magnetic field.

Magnetic fields are typically well-ordered across
large spatial scales, ranging from approximately 100
parsecs to 1 parsec \citep{2000AJ....119..923H}. However, on smaller, sub-parsec scales, magnetic fields can exhibit random orientations. This disarray is attributed to processes such as ambipolar diffusion \citep{1956MNRAS.116..503M, 1993ApJ...415..680F, 2009MNRAS.399.1681T, masson2016ambipolar} and turbulent magnetic reconnection diffusion \citep{2005AIPC..784.....D, 2013ApJ...777...46L}, which disrupt the coherence of the magnetic field lines.

In the paper, we used the JCMT/SCUPOL data from legacy surveys in some cores and filaments and the TADPOL archival data in some regions which are common in both studies. We used the available polarization measurements to infer: mean B-field orientations, the statistical distribution of polarization values, relations between large- (JCMT at 15" resolution) and small- (CARMA at 2.5" resolution) scale B-fields, and the relation between outflow orientations with mean B-field orientations at different spatial scales. This work uses the opportunity to do statistical analysis of the available polarization measurements which was not presented in the original paper of JCMT/SCUPOL legacy survey \citep{2009ApJS..182..143M}. Here section 2 presents data acquisition details, section 3 shows the results of our analysis, and section 4 summarises this study.

\section{Methodology}

\subsection{The JCMT-SCUPOL data}

In this study, we utilized SCUPOL data to explore the statistical properties of magnetic fields within core and filamentary structures. This dataset, derived from the comprehensive work \textit{The Legacy of SCUPOL: 850 $\mu$m Imaging Polarimetry from 1997 to 2005}, encompasses polarimetric observations performed by SCUPOL, the polarimeter for SCUBA on the James Clerk Maxwell Telescope (JCMT), covering the period from 1997 to July 2005 \citep{2009ApJS..182..143M}. While SCUBA facilitated simultaneous observations at both 850 $\mu$m and 450 $\mu$m, the 450 $\mu$m data were either unavailable or had insufficient sensitivity for the regions analyzed. As a result, our study is exclusively based on 850 $\mu$m observations, targeting various astrophysical environments, including cores, filaments, ridges, galaxies, planetary nebulae, asymptotic giant branch (AGB) stars, and supernova remnants, with a particular emphasis on cores and filaments. Within core regions, we focused on four primary categories: Bok globules (BG), starless or pre-stellar cores (SC/PC), star-forming regions (SFR), and young stellar objects (YSO). Data reduction and processing were accomplished using the Starlink software suite, including the SURF, KAPPA, POLPACK, and CURSA packages, yielding an angular resolution of 15" and the detailed data reduction procedure is described in \citep{2009ApJS..182..143M}. Further, we supplemented our analysis with data from additional specific cores of interest.

\subsection{The CARMA data}

To examine the dynamic interplay between outflows from core regions and magnetic fields across different spatial scales, we incorporated data from the TADPOL survey, as detailed in the publication \textit{TADPOL: A 1.3 mm Survey of Dust Polarization in Star-Forming Cores and Regions} \citep{2014ApJS..213...13H}. This survey presents 1.3 mm polarimetric observations from the Combined Array for Research in Millimeter-wave Astronomy (CARMA), encompassing dust polarization data for 30 star-forming cores and 8 star-forming regions. The TADPOL data provide high-resolution magnetic field maps on a more localized scale, with a spatial resolution of 2.5", offering a finer level of detail than the JCMT/SCUPOL observations, thereby enhancing our ability to probe magnetic field structure on smaller scales. The data reduction methodology for the CARMA observations is thoroughly described in \cite{2014ApJS..213...13H}.

\section{Results}
\subsection{Statistical analysis of the core polarization data}

\subsubsection{Magnetic fields in the cores:}  

To examine the morphology of magnetic fields within core regions, we selected a subset of 45 cores from the dataset presented by \cite{2009ApJS..182..143M}, which involved dust polarization observations conducted at a wavelength of 850 $\mu$m using the JCMT/SCUPOL system, achieving a resolution of 15". In this study, data were sampled at an angular resolution of 10"—with select regions binned to 20", as elaborated in \cite{2009ApJS..182..143M}—and included only areas exhibiting significant polarization vectors. The selected regions meet the stringent criteria of \( p/dp > 2 \), \( dp < 4\% \), and \( I > 0 \), where \( I \) represents the Stokes intensity parameter and \( p \) denotes the polarization degree.

To investigate the variation in outflow direction in relation to the mean magnetic field direction across different scales, we further narrowed our focus to nine core regions shared between the 30 star-forming cores and 8 star-forming regions observed in the Combined Array for Research in Millimeter-wave Astronomy (CARMA) at a wavelength of 1.3 mm with a resolution of 2.5" \citep{2014ApJS..213...13H}. These nine core regions are common to both the CARMA and JCMT observations, allowing for a comparative analysis of magnetic field dynamics in these astrophysical environments.

\begin{figure*}[h]
\centering
\includegraphics[width=1.9\columnwidth]{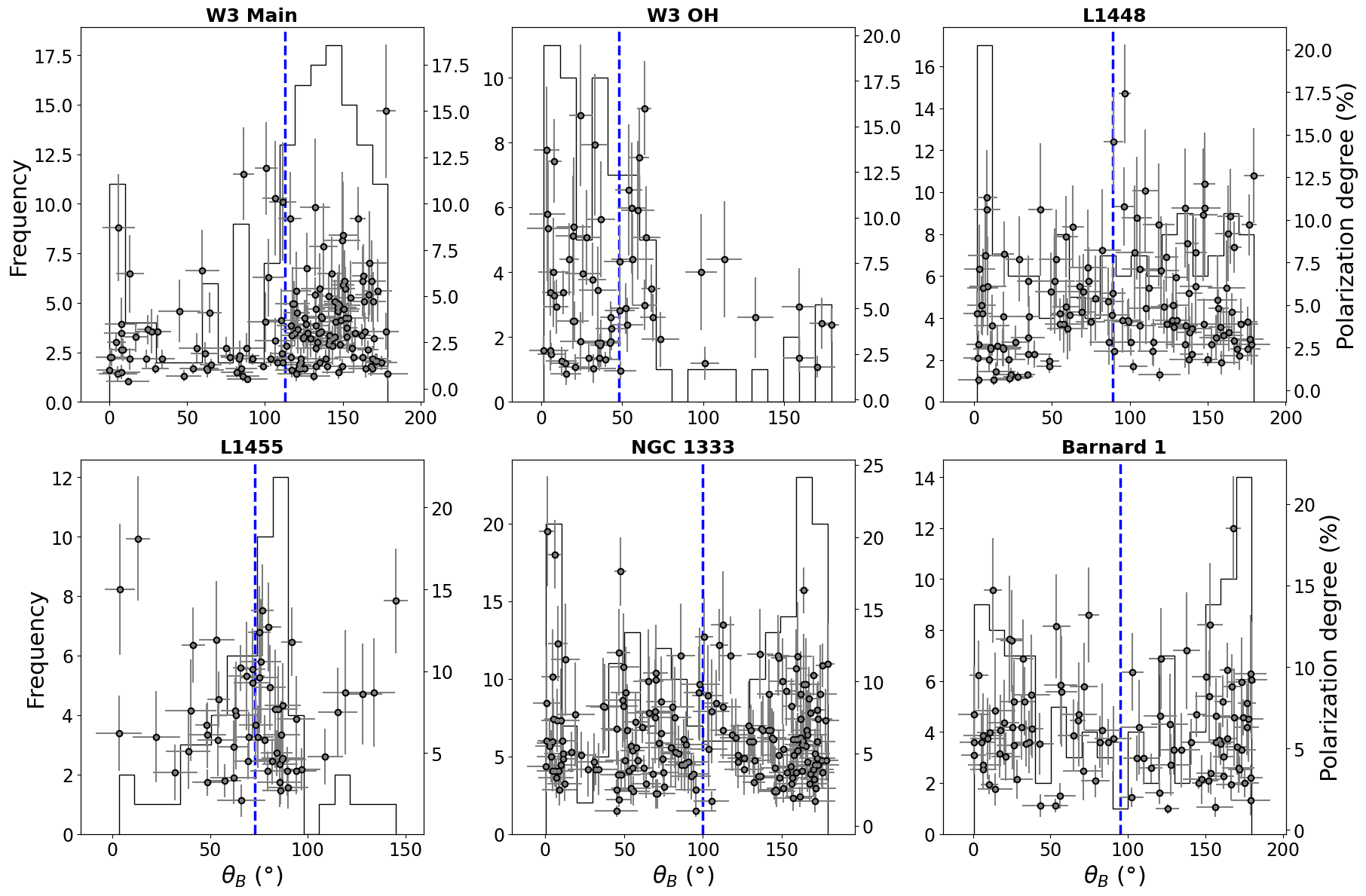}
\caption{Histogram of $\theta_B$ and scatter plot of polarization degree (p) vs. $\theta_B$ with error bars for the core regions. The histograms illustrate the distribution of magnetic field vector position angles ($\theta_B$), with a blue dotted line in each plot indicating the mean position angle ($\theta_{B_\text{mean}}$) for the respective core regions. The scatter plots depict the polarization degree (p) across various position angles in these regions, with error bars representing the uncertainties associated with each value of p and $\theta_B$. }\label{Histogram and scatter plot of core}
\end{figure*}

\begin{table*}[htb]
\centering
\scriptsize 
    \caption{Table for calculating $\Delta\theta$ (= $|\theta_{B_{mean}} - \theta_{minor}|$)}\label{tableExample} 
    \begin{tabular}{lccccccc}
         \hline
         Name & RA (J2000) & DEC (J2000) & Object type & $\theta_{minor}$($^{\circ}$) & $\theta_{B_{mean}}$($^{\circ}$) & $\Delta\theta$($^{\circ}$) & Distance (kpc) \\
         \hline
         W3 Main & 02 25 35.44 & +62 06 16.4 & SFR (HM) & 17 & $113 \pm 9$ & $96 \pm 9$ & 1.95 \citep{xu2006distance} \\ 
         W3 OH & 02 27 03.83 & +61 52 24.8 & SFR (HM) & 141 & $48 \pm 9$ & $93 \pm 9$ & 1.95 \citep{xu2006distance} \\
         L1448 & {03 25 38.80} & {+30 44 05.4} & {YSO (Class 0)} & 80 & $89 \pm {10}$ & $9 \pm {10}$ & {$0.250 \pm 0.050$} \citep{enoch2006bolocam} \\
         L1455 & {03 27 41.31} & {+30 12 39.4} & {SFR (LM)} & 45 & $73 \pm {9}$ & $28 \pm {9}$ & {$0.250 \pm 0.050$} \citep{enoch2006bolocam} \\
         NGC 1333 & {03 29 11.11} & {+31 13 20.2} & {SFR (HM)} & 89 & $100 \pm {9}$ & $11 \pm {9}$ & {0.320} \citep{de1999hipparcos} \\
         Barnard 1 & {03 33 17.88} & {+31 09 32.9} & {SFR (LM)} & 99 & $95 \pm {9}$ & $4 \pm {9}$ & {$0.250 \pm 0.050$} \citep{enoch2006bolocam} \\
         HH211/IC348 & {03 43 56.70} & {+32 00 51.9} & {SFR (LM)} & 120 & $114 \pm {6}$ & $6 \pm {6}$ & {$0.250 \pm 0.050$} \citep{enoch2006bolocam} \\
         L1498 & {04 10 52.60} & {+25 10 00.0} & {SC/PC} & 36 & $123 \pm {9}$ & $87 \pm {9}$ & {$0.14 \pm 0.02$} \citep{ungerechts1987co} \\
         L1527 & {04 39 53.90} & {+26 03 10.0} & {YSO (Class 0)} & 35 & $75 \pm {9}$ & $40 \pm {9}$ & {$0.140 \pm 0.010$} \citep{kenyon1994new} \\
         IRAM 04191+1522 & {04 21 56.91} & {+15 29 46.1} & {YSO (Class 0)} & 164 & $64 \pm {8}$ & $100 \pm {8}$ & {$0.140 \pm 0.010$} \citep{kenyon1994new} \\
         L1544 & {05 04 17.23} & {+25 10 43.7} & {SC/PC} & 55 & $61 \pm {10}$ & $6 \pm {10}$ & {$0.140 \pm 0.010$} \citep{kenyon1994new} \\
        NGC 2071 IR & {05 47 04.85} & {+00 21 47.1} & {SFR (HM)} & 80 & $89 \pm {10}$ & $9 \pm {10}$ & {0.400} \citep{anthony1982h} \\
         IRAS 05490+2658 & {05 52 13.24} & {+26 59 33.3} & {SFR (HM)} & 77 & $71 \pm {10}$ & $6 \pm {10}$ & {2.1} \citep{snell1990molecular} \\
         Mon R2 IRS & {06 07 46.16} & {-06 23 22.5} & {SFR (HM)} & 110 & $51 \pm {8}$ & $59 \pm {8}$ & {0.950} \citep{racine1970reflection} \\
         MON IRAS 12 & {06 41 05.81} & {+09 34 09.0} & {SFR (HM)} & 115 & $67 \pm {5}$ & $48 \pm {5}$ & {0.800} \citep{1956ApJS....2..365W} \\
         CB 54 & {07 04 21.07} & {-16 23 20.09} & {BG} & 70 & $78 \pm {10}$ & $8 \pm {10}$ & {1.1} \citep{brand1993velocity} \\
         L183 & {15 54 08.96} & {-02 52 43.9} & {SC/PC} & 80 & $86 \pm {9}$ & $6 \pm {9}$ & {0.15} \citep{ward1999initial} \\
         $\rho$ Oph A & {16 26 26.45} & {-24 24 10.9} & {SFR (IM)} & 96 & $103 \pm {8}$ & $7 \pm {8}$ & {0.139} \citep{mamajek2008distance} \\
         $\rho$ Oph C & {16 27 00.10} & {-24 34 26.7} & {SFR (IM)} & 25 & $88 \pm {9}$ & $63 \pm {9}$ & {0.139} \citep{mamajek2008distance} \\
         $\rho$ Oph B2 & {16 27 27.97} & {-24 27 06.8} & {SFR (IM)} & 170 & $93 \pm {9}$ & $77 \pm {9}$ & {0.139} \citep{mamajek2008distance} \\
         L43 & {16 34 35.57} & {-15 47 00.6} & {SC/PC} & 39 & $101 \pm {9}$ & $62 \pm {9}$ & {0.17} \citep{ward1999initial} \\
         NGC 6334A & {17 20 19.55} & {-35 54 42.3} & {SFR (HM)} & 158 & $97 \pm {10}$ & $61 \pm {10}$ & {1.7} \citep{neckel1978ubv} \\
         GGD 27 & {18 19 12.00} & {-20 47 30.9} & {SFR (HM)} & 117 & $74 \pm {10}$ & $43 \pm {10}$ & {1.7} \citep{rodriguez1980radio} \\
         CRL 2136 IRS 1 & {18 22 26.48} & {-13 30 15.1} & {SFR (HM)} & 100 & $106 \pm {10}$ & $6 \pm {10}$ & {2} \citep{kastner1992juggler} \\
         Serpens Main Core & {18 29 49.34} & {+01 15 54.6} & {SFR (LM)} & 65 & $97 \pm {8}$ & $32 \pm {8}$ & {0.310} \citep{de1991distance} \\
         CL 04/CL 21 & {18 37 19.39} & {-07 11 31.8} & {SFR (HM)} & 108 & $119 \pm {9}$ & $11 \pm {9}$ & {0.770} \citep{webster1976some} \\
         G28.34+0.06 & {18 42 52.40} & {-03 59 53.9} & {SFR (HM)} & 89 & $104 \pm {10}$ & $15 \pm {10}$ & {4.8} \citep{carey2000submillimeter} \\
         IRAS 18437-0216 & {18 46 23.23} & {-02 13 45.4} & {SFR (HM)} & 84 & $77 \pm {9}$ & $7 \pm {9}$ & {6.6} \citep{sridharan2005high} \\
         W48 & {19 01 45.45} & {+01 13 04.5} & {SFR (HM)} & 10 & $76 \pm {8}$ & $66 \pm {8}$ & {3.4} \citep{vallee1990co} \\
         R Cr A & {19 01 53.65} & {-36 57 07.5} & {SFR (HM)} & 99 & $100 \pm {10}$ & $1 \pm {10}$ & {0.130} \citep{marraco1981distance} \\
         W49 & {19 10 13.60} & {+09 06 17.4} & {SFR (HM)} & 100 & $100 \pm {6}$ & $0 \pm {6}$ & {11.4} \citep{gwinn1992distance} \\
         W51 & {19 23 42.00} & {+14 30 33.0} & {SFR (HM)} & 90 & $98 \pm {9}$ & $8 \pm {9}$ & {7.5} \citep{genzel1981proper} \\
         IRAS 20081+2720 & {20 10 13.99} & {+27 28 36.9} & {SFR (LM)} & 14 & $29 \pm {9}$ & $15 \pm {9}$ & {0.700} \citep{wilking1989millimeter} \\
         AFGL 2591 IRS & {20 29 24.72} & {+40 11 18.9} & {SFR (HM)} & 150 & $114 \pm {9}$ & $36 \pm {9}$ & {1.5} \citep{wendker1974radio} \\
         IRAS 20188+3928 & {20 20 38.75} & {+39 38 03.9} & {SFR (HM)} & 98 & $110 \pm {9}$ & $12 \pm {9}$ & {0.4 – 4} \citep{little1988iras} \\
         S106 & {20 27 17.32} & {+37 22 41.3} & {SFR (HM)} & 155 & $96 \pm {8}$ & $59 \pm {8}$ & {0.600} \citep{staude1982bipolar} \\
         G079.3+0.3 & {20 32 23.62} & {+40 19 44.0} & {SFR (LM)} & 117 & $130 \pm {9}$ & $13 \pm {9}$ & {1} \citep{carey2000submillimeter} \\
         S140 & {22 19 18.00} & {+63 18 49.0} & {SFR (HM)} & 92 & $87 \pm {8}$ & $5 \pm {8}$ & {0.900} \citep{preibisch2002outflow} \\
         S146 & {22 49 28.56} & {+59 55 08.6} & {SFR (HM)} & 89 & $89 \pm {9}$ & $0 \pm {9}$ & {5.2} \citep{wu2005co} \\
         Cepheus A & {22 56 17.80} & {+62 01 49.0} & {YSO (Class 0/I)} & 147 & $71 \pm {8}$ & $76 \pm {8}$ & {0.730} \citep{blaauw1959photoelectric} \\
         S152 & {22 58 50.14} & {+58 45 01.0} & {SFR (HM)} & 71 & $122 \pm {9}$ & $51 \pm {9}$ & {5} \citep{wouterloot1993iras} \\
         NGC 7538 & {23 13 45.54} & {+61 27 35.7} & {SFR (HM)} & 90 & $113 \pm {8}$ & $23 \pm {8}$ & {2.8} \citep{blitz1982catalog} \\
         S157 & {23 16 04.00} & {+60 02 06.0} & {SFR (HM)} & 64 & $79 \pm {10}$ & $15 \pm {10}$ & {2.5} \citep{shirley2003cs} \\
         \hline
    \end{tabular}
    \begin{tablenotes}
        \small
       The values for $\theta_{B_{mean}}$ presented in this table are derived from JCMT/SCUPOL observations, while the classifications for "Object Type", "Distance", and the locations of the regions (RA and DEC) are adopted from \cite{2009ApJS..182..143M}. Right Ascension (RA) and Declination (DEC) are specified in the conventional units of hours, minutes, and seconds for RA, and degrees, arcminutes, and arcseconds for DEC. The references for the distances associated with the regions taken from \cite{2009ApJS..182..143M} are presented in parentheses adjacent to the corresponding values. The star-forming regions (SFRs) are categorized by mass as high-mass (HM), low-mass (LM), and intermediate-mass (IM), with these designations indicated in parentheses following the respective SFRs.
    \end{tablenotes}
\end{table*}

\subsubsection{Histogram of $\theta_B$ and scatter plot of p vs. $\theta_B$:}
In the JCMT/SCUPOL dataset, the position angles (\(\theta_E\)) of the polarization E-vector at a wavelength of 850 $\mu$m are provided. To derive the position angle (\(\theta_B\)) of the B-vector, \(\theta_E\) is rotated by 90°, after which a histogram is constructed to illustrate the distribution of \(\theta_B\) for each individual core region. The blue dotted line in these histograms represents the mean position angle of the magnetic fields (\(\theta_{B_{\text{mean}}}\)) across the respective core regions.

Additionally, the scatter plot depicting the degree of polarization (p) against the B-vector position angle (\(\theta_B\)) reveals the relationship between polarization and magnetic field orientation. Figure 1
displays the histogram of \(\theta_B\) alongside the scatter plot of p versus \(\theta_B\) for selected cores. The error bars in the scatter plots indicate the uncertainties associated with each p and \(\theta_B\) measurement.

The scatter plots, inclusive of error bars, reveal that the majority of data points cluster within the ranges of \(113^\circ < \theta_B < 175^\circ\) and \(0.8\% < p < 6\%\) for the W3 Main region, as well as within \(0^\circ < \theta_B < 75^\circ\) and \(1\% < p < 16\%\) for the W3 OH region. In contrast, the remaining plots exhibit a greater dispersion in both p and \(\theta_B\) values, indicating a more complex relationship in those regions.

\vspace{0.3em}

\begin{figure*}[h]
\centering
\includegraphics[width=1.9\columnwidth]{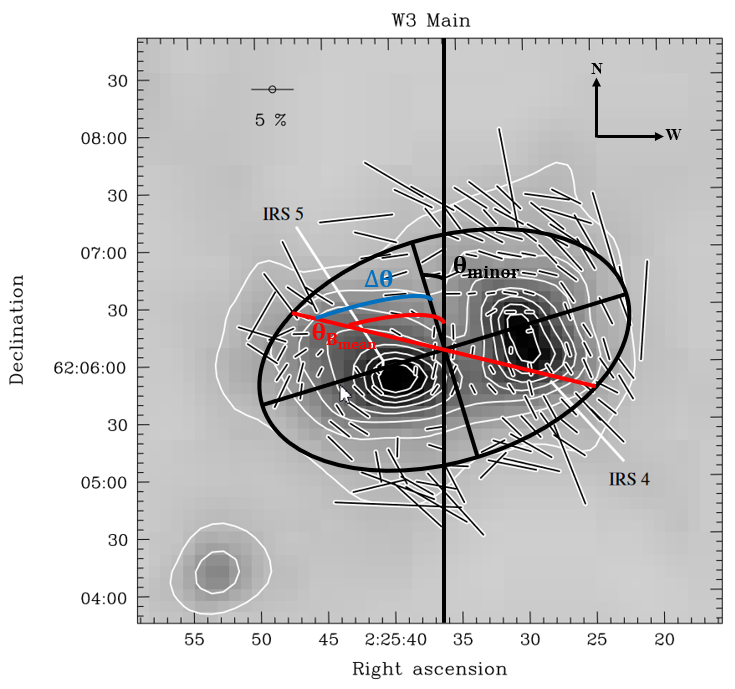}
\caption{Measurement of the minor axis angle ($\theta_\text{minor}$), the mean magnetic field angle ($\theta_{B_{\text{mean}}}$), and the angular difference between these orientations ($\Delta \theta$) for the W3 Main region. The color map is adapted from \cite{2009ApJS..182..143M}, as the polarization map was not re-created in this study.}\label{Measurement of theta minor for core}
\end{figure*}

\subsubsection{Histogram and Kernal Density Estimation (KDE) of the difference between $\theta_{minor}$ and $\theta_{B_{mean}}$}:

The angle (\(\theta_{\text{minor}}\)) formed by the minor axis of an ellipse, which is roughly fitted to the core region, with respect to the vertical line (the north-south direction) is measured in an anticlockwise manner. This ellipse is approximated either inside or outside the outermost contour of the core region, as delineated in \cite{2009ApJS..182..143M}. The B-vector position angles (\(\theta_B\)) for individual cores are derived by adding \(90^\circ\) to the \(\theta_E\) values provided in the SCUPOL dataset. Subsequently, the mean B-vector position angle (\(\theta_{B_{\text{mean}}}\)) is computed for each core. Figure 2 illustrates the measurement procedure for \(\theta_{\text{minor}}\), \(\theta_{B_{\text{mean}}}\), and \(\Delta \theta\) with respect to the reference direction (north-south).

The analysis summarized in Table 1 reveals a striking alignment in certain cores, notably W49 and S146, where the \(\Delta\theta\) value approaches \(0^\circ\), defined as \(\Delta\theta = |\theta_{B_{\text{mean}}} - \theta_{\text{minor}}|\). This close alignment signifies that the mean magnetic field in these cores is precisely oriented along the minor axis. In contrast, this parallelism is less pronounced or absent in other cores.

To illustrate the distribution of \(\Delta\theta\), we generated a histogram, displayed in Figure 3, with a red curve representing a Gaussian fit. This fit peaks at approximately \(6.70^\circ\), indicating a prevalent trend in \(\Delta\theta\) alignment. Figure 4 presents the Kernel Density Estimate (KDE) plot of the \(\Delta\theta\) histogram, peaking at around \(7.21^\circ\), reinforcing the observed alignment trend. Both the Gaussian fit and KDE plot peaks suggest that the majority of cores exhibit a minor axis orientation closely aligned with the mean magnetic field direction.

From Table 1, we calculate the mean \(\Delta\theta\) value as \(32^\circ \pm 9^\circ\). This average supports the notion that the magnetic field direction within these core regions predominantly aligns with the minor axis. Such an orientation is consistent with observations by SCUPOL, highlighting a significant degree of parallelism between the magnetic field and structural morphology within these astrophysical cores. Additionally, \cite{soam2015magnetic} showed that the mean magnetic field direction with the minor axes of a limited sample of five cores is approximately \(37^\circ\), further implying that the mean B-field is nearly parallel to the minor axis of these core regions.

The dispersion of \(\Delta \theta\) values within the core regions is approximately \(30.19^\circ\). This substantial dispersion indicates that the B-field position angles exhibit a random alignment in the JCMT-SCUPOL observations. Overall, this study provides a statistical representation of the magnetic field morphology in the plane of the sky, consistent with previous hypotheses.

\begin{figure*}[h]
\centering
\includegraphics[width=\linewidth]{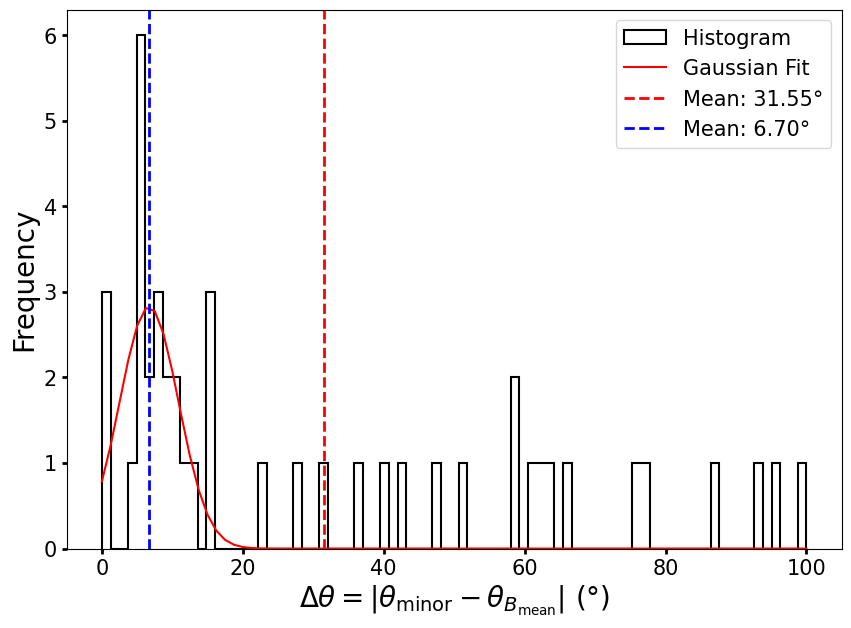}
\caption{Histogram of the angular difference, $\Delta\theta$ ($|\theta_{B_\text{mean}} - \theta_\text{minor}|$), for the core regions. This histogram illustrates the distribution of $\Delta\theta$ values computed for the cores identified in Table 1, with a red curve superimposed to depict the Gaussian fit to the data. The vertical dashed lines, colored red and blue, represent the mean value of $|\Delta\theta|$ and the peak of the fitted Gaussian curve, respectively, providing insight into the central tendency and most probable value of the angular difference.} \label{Histogram for delta theta}
\end{figure*}

\begin{figure*}[h]
\centering
\includegraphics[width=\linewidth]{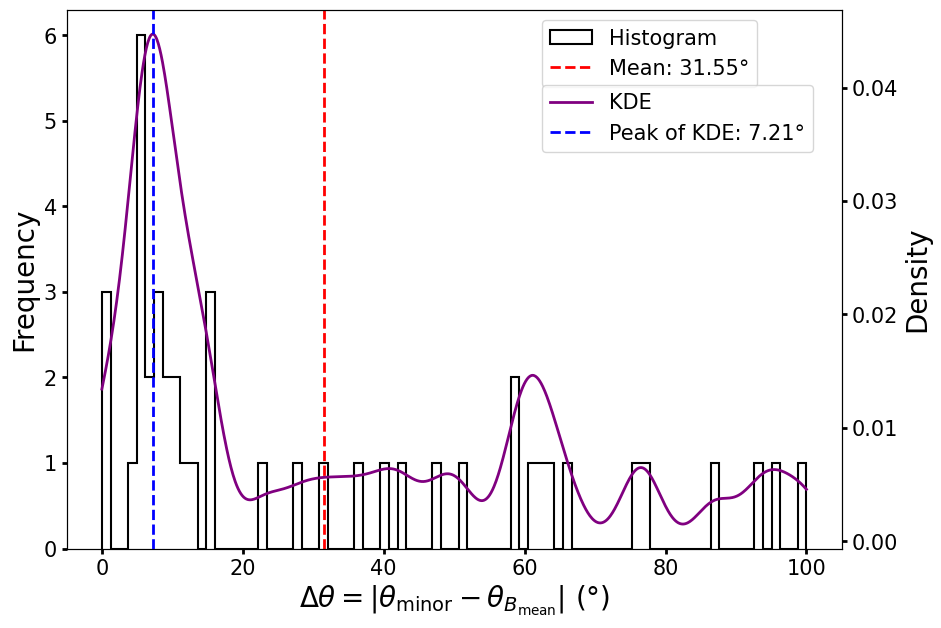}
\caption{KDE plot (depicted by the purple curve) of the angular difference, $\Delta\theta$ (= $|\theta_{B_\text{mean}} - \theta_\text{minor}|$), for the core regions, overlaid on the histogram of $|\Delta\theta|$ values derived from the cores listed in Table 1. The red and blue dashed lines indicate the mean value of $\Delta\theta$ and the peak of the KDE curve, respectively, highlighting the central tendency and the most probable value of the angular difference within the data distribution.}\label{KDE plot for theta minor and theta B mean}
\end{figure*}

\vspace{0.2em}

\subsubsection{Difference between outflow and mean B-field orientation:}

In the TADPOL survey \citep{2014ApJS..213...13H}, a detailed examination of dust polarization was conducted using observations from the Combined Array for Research in Millimeter-wave Astronomy (CARMA) at a wavelength of 1.3 mm, targeting 30 star-forming cores and 8 star-forming regions. Through these observations, the small-scale magnetic field orientation (\(\chi_{\text{sm}}\)) was determined with a resolution of \(2''.5\) within the identified core regions. Additionally, the angle of the outflow ejection (\(\chi_0\)) was also calculated.

Prior to section 3.1.4, the mean magnetic field position angle (\(\theta_{B_{\text{mean}}}\)) for each individual core region was derived from JCMT/SCUPOL data. Notably, \(\theta_{B_{\text{mean}}}\) represents the mean value of the B-vector position angle on a large scale, which is significantly larger than the scale of the CARMA observations.

All angles—\(\chi_{\text{sm}}\), \(\chi_0\), and \(\theta_{B_{\text{mean}}}\)—are measured in an anticlockwise direction with respect to the north-south orientation.

Between the CARMA and JCMT surveys, we identified nine common core regions that facilitated a comparative analysis of the variation in outflow direction between large-scale and small-scale magnetic fields, which we visualized using a Kernel Density Estimate (KDE) plot.

The KDE plot depicting the angular difference between \(\theta_{B_{\text{mean}}}\) and \(\chi_0\) is shown in Figure 5. In this plot, the blue dashed line highlights a peak value of \(|\theta_{B_{\text{mean}}} - \chi_0|\) around \(17.25^\circ\), while the mean value, calculated from Table 2, is approximately \(31^\circ \pm 8^\circ\). These findings collectively suggest that the outflow emanating from the core regions aligns closely with the mean magnetic field direction on large scales. In a study by \cite{soam2015magnetic}, an examination of five cores containing Very Low Luminosity Objects (VeLLOs) showed that the outflows in three of these cores exhibit a significant tendency to align with the magnetic field orientation within the surrounding envelope.

Following this, we present the KDE plot for the angular difference between \(\chi_0\) and \(\chi_{\text{sm}}\) in Figure 6. Here, the blue dashed line indicates a peak value of \(|\chi_{\text{sm}} - \chi_0|\) at \(74.87^\circ\), while the mean value, derived from Table 2, is approximately \(70^\circ \pm 36^\circ\). These results imply that, at smaller scales, the outflow direction within the core regions is predominantly orthogonal to the mean magnetic field orientation. This finding is consistent with \cite{2014ApJS..213...13H}, who observed that, on smaller scales, the outflow direction tends to be nearly perpendicular to the mean magnetic field position angle.

Table 2 consolidates the computed values of \(|\theta_{B_{\text{mean}}} - \chi_0|\) and \(|\chi_{\text{sm}} - \chi_0|\) across the nine cores examined, providing a clear summary of the observed alignment trends at both large and small scales.

\begin{figure*}[h]
\centering
\includegraphics[width=\linewidth]{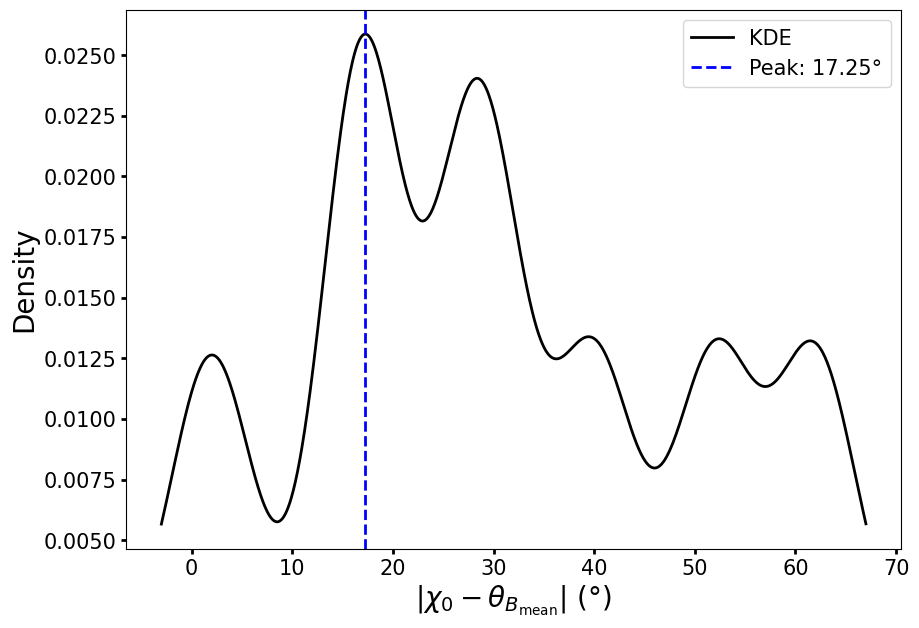}
\caption{KDE plot depicting the angular difference between $\theta_{B_\text{mean}}$ and $\chi_0$ for the nine cores observed in both JCMT and CARMA datasets. The blue dashed line marks the peak of the KDE curve for $|\theta_{B_\text{mean}} - \chi_0|$, which occurs at $17.25^\circ$. This value suggests that, on a large scale, the outflow direction for these cores is nearly aligned with the mean magnetic field orientation.} \label{KDE plot for theta B mean and chi 0}
\end{figure*}

\begin{figure*}[h]
\centering
\includegraphics[width=\linewidth]{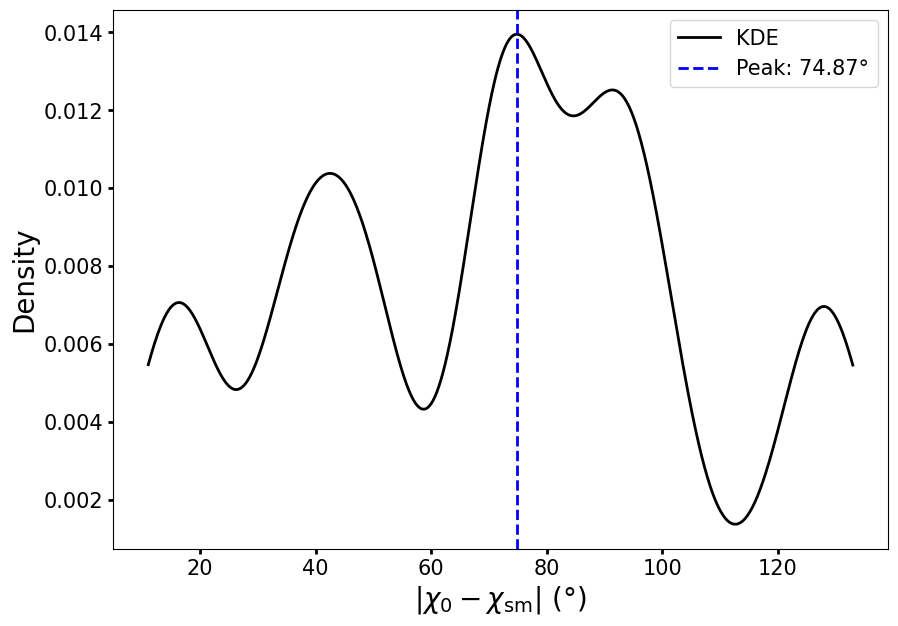}
\caption{KDE plot depicting the angular difference between $\chi_{sm}$ and $\chi_0$ for the nine cores observed in both JCMT and CARMA datasets. The blue dashed line marks the peak of the KDE curve for $|\chi_0 - \chi_{sm}|$, which occurs at $74.87^\circ$. This value suggests that, on a small scale, the outflow direction for these cores is nearly perpendicular to the mean magnetic field orientation.} \label{KDE plot for chi 0 and chi sm}
\end{figure*}

\begin{table*}[htb]
\centering
\scriptsize
    \caption{Table for observational values of small-scale magnetic field $\chi_{sm}$ and outflow direction $\chi_0$ for core regions}\label{tableExample} 
    \begin{tabular}{lccccccccccc}
    \hline
    No. & Name & {p (\%)} & $\chi_{sm}$ ($^{\circ}$) & $\chi_0$ ($^{\circ}$) & $\theta _{B_{mean}}$ ($^{\circ}$) & $|\theta_{B_{mean}} - \chi_0|$ ($^{\circ}$) & $|\chi_{sm} - \chi_0|$ ($^{\circ}$)\\\hline
    \multirow{3}{*}{1} & L1448N & {$5 \pm 2$} & $26 \pm {37}$ & 97& $89 \pm {10}$ & $40 \pm {10}$ & $71 \pm {35}$ &\\ & L1448C & {$5 \pm 2$} & $112 \pm {32}$ & 161 & $89 \pm {10}$ & $40 \pm {10}$ & $71 \pm {35}$\\
    \hline
    \multirow{3}{*}{2} & NGC 1333- & {$6 \pm 2$} & $70 \pm {23}$ & 59.5& $100 \pm {9}$ & $62 \pm {9}$ & $16 \pm {25}$ &\\ & IRAS 2$A^c$\\ & SVS 13 & {$6 \pm 2$} &$6 \pm {24}$ & ... & $100 \pm {9}$ & $62 \pm {9}$ & $16 \pm {25}$ &\\ & NGC 1333- & {$6 \pm 2$} & $56 \pm {20}$ & 18& $100 \pm {9}$ & $62 \pm {9}$ & $16 \pm {25}$ &\\ & IRAS 4A\\ & NGC 1333- & {$6 \pm 2$} & $84 \pm {34}$ & 0 & $100 \pm {9}$ & $62 \pm {9}$ & $16 \pm {25}$ &\\ & IRAS 4B\\ & NGC 1333- & {$6 \pm 2$} & $55 \pm {20}$ & 76 & $100 \pm {9}$ & $62 \pm {9}$ & $16 \pm {25}$\\ & IRAS-4B2\\
    \hline
    3 & HH211MM & {$9 \pm 2$} & $164 \pm {32}$ & 116 & $114 \pm {6}$ & $2 \pm {6}$ & $48 \pm {32}$\\
    \hline
    4 & L1527 & {$7 \pm 2$} & $3 \pm {8}$ & 92 & $75 \pm {9}$ & $17 \pm {9}$ & $89 \pm {8}$\\
    \hline
    5 & CB 54 & {$6 \pm 2$} & $32 \pm {42}$ & 108 & $78 \pm {10}$ & $30 \pm {10}$ & $76 \pm {42}$\\
    \hline
    6 & VLA 1623 & {$5 \pm 1$} & $23 \pm {48}$ & 120 & $103 \pm {8}$ & $17 \pm {8}$ & $97 \pm {48}$\\
    \hline
    \multirow{3}{*}{7} & Ser-emb 8 & {$7 \pm 2$} & $7 \pm {44}$ & 123 & $97 \pm {8}$ & $27 \pm {8}$ & $37 \pm {33}$ &\\ & Ser-emb 8(N) & {$7 \pm 2$} & $83 \pm {15}$ & 107 & $97 \pm {8}$ & $27 \pm {8}$ & $37 \pm {33}$ &\\ & Ser-emb 6 & {$7 \pm 2$} & $172 \pm {33}$ & 135 & $97 \pm {8}$ & $27 \pm {8}$ & $37 \pm {33}$\\
    \hline
    8 & NGC 7538 & {$4 \pm 1$} & $52 \pm {62}$ & ... & $113 \pm {8}$ & ... & ... & \\
    \hline
    9 & CB 244 & {$12 \pm 3$} & $170 \pm {49}$ & 42 & $94 \pm {7}$ & $52 \pm {7}$ & $128 \pm {49}$\\
    \hline
    \end{tabular}
    \tablenotes{The polarization degree ($p$) and mean magnetic field angle, $\theta_{B_{\text{mean}}}$, are derived from JCMT-SCUPOL data \citep{2009ApJS..182..143M}. The outflow orientation, $\chi_0$, along with the direction of the small-scale magnetic field, $\chi_{\text{sm}}$, are adopted from the comprehensive analysis presented in \cite{2014ApJS..213...13H}.
    }
\end{table*}

\subsubsection{Kernel Density Estimation plot of $\chi_{sm}$ (CARMA data) and $\theta_{B_{mean}}
$ (JCMT data):}

This study demonstrated that the alignment between the large-scale and small-scale magnetic fields in core regions was not entirely parallel, as illustrated in Figure 7. The small-scale observations, which provided a magnified view of the larger scales, revealed a more intricate structure of the magnetic field orientation deep within the cores.

This observed misalignment between the large-scale and small-scale magnetic field directions contributed to a low degree of polarization (\(p\)). This phenomenon suggested that the alignment of dust grains was disrupted by the random orientations of magnetic fields at smaller scales \citep{2014ApJS..213...13H}. The transition from large to small spatial scales unveiled a significantly more complex polarization pattern, characterized by distorted magnetic field geometries. Furthermore, evidence of hourglass morphologies was apparent in the densest regions of some cores, indicating the intricate interplay between the magnetic fields and the surrounding matter.

\begin{figure*}[h]
\centering
\includegraphics[width=\linewidth]{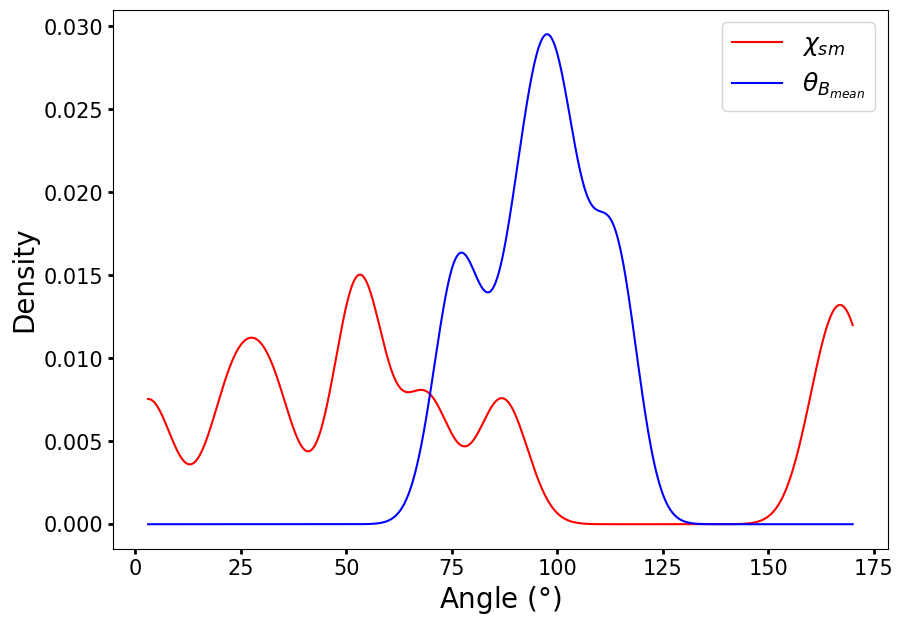}
\caption{KDE plot of $\chi_{sm}$ and $\theta_{B_{mean}}$ for nine cores common in CARMA and JCMT observations which clearly shows the misalignment of large and small scale mean magnetic field direction in those nine core regions.}\label{KDE plot chi sm and theta b mean}
\end{figure*}

\begin{figure*}[h]
\centering
\includegraphics[width=1.92\columnwidth]{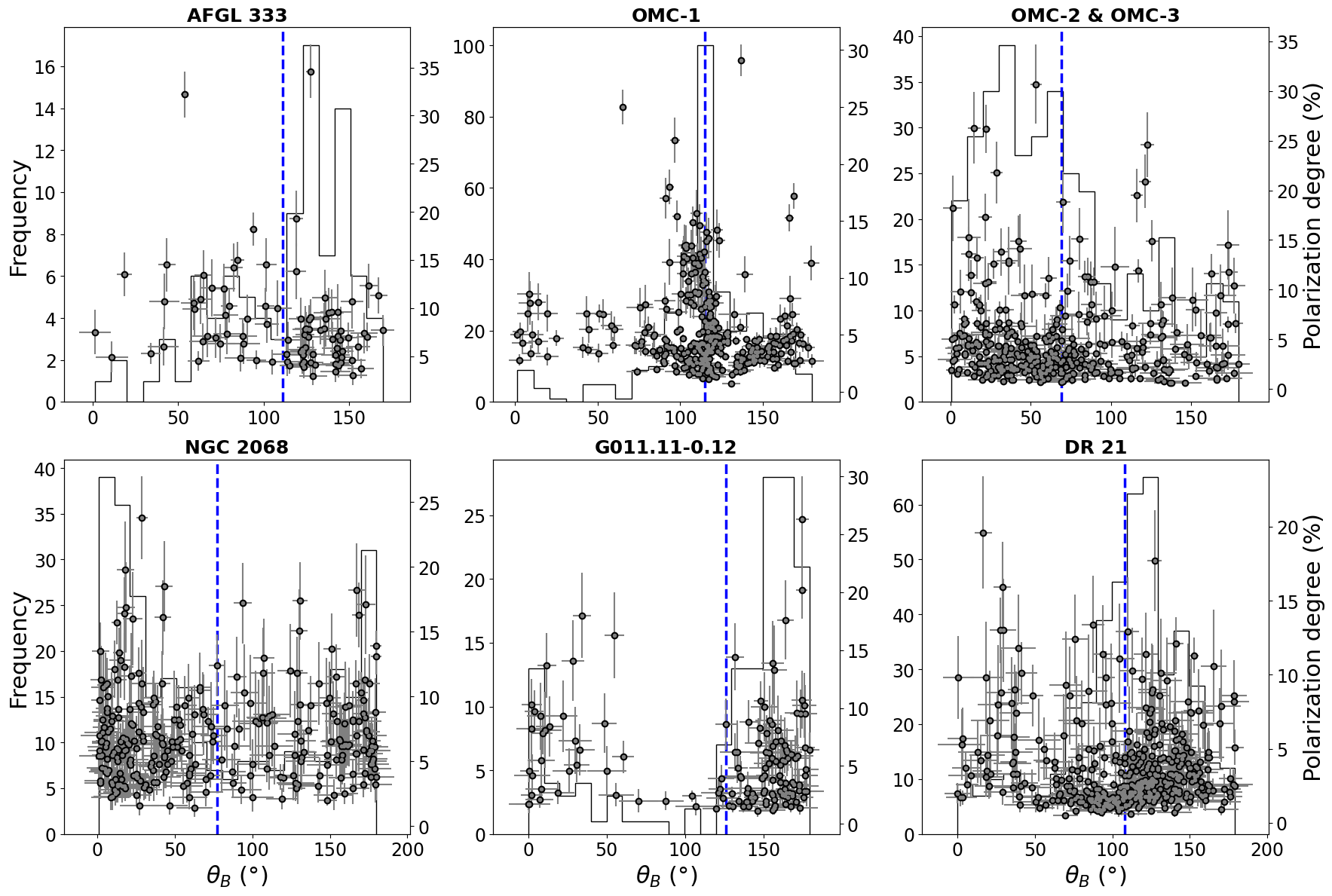}
\caption{Histogram of $\theta_B$ and scatter plot of p vs. $\theta_B$ with the error bars for the filament regions. The histograms show the distribution of B-vector position angles ($\theta_B$), and the blue dotted line in each of the histograms shows the mean value of B-vector position angle ($\theta_{B_{mean}}$) in those individual filament regions. The scatter plots show the polarization degree (p) at different position angles for those filament regions and the error bars are the uncertainties in each p and $\theta_B$ value.}\label{Histogram and scatter plot for filament region}
\end{figure*}

\begin{figure*}[h]
\centering
\includegraphics[width=1.9\columnwidth]{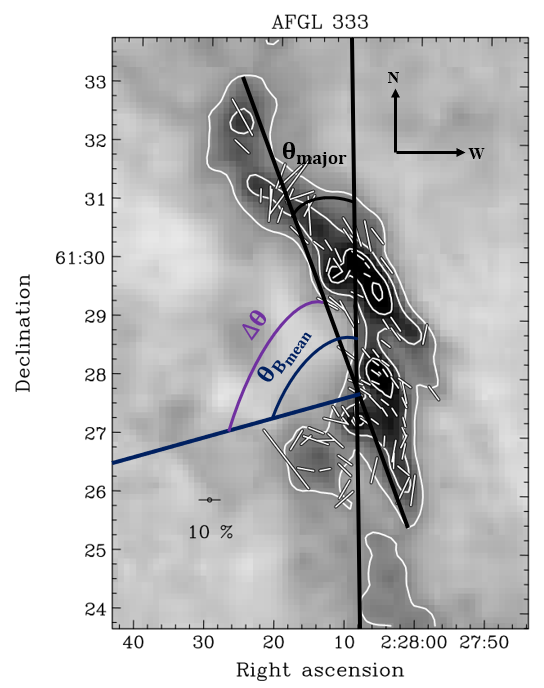}
\caption{Measurement of the minor axis angle ($\theta_\text{major}$), the mean magnetic field angle ($\theta_{B_{\text{mean}}}$), and the angular difference between these orientations ($\Delta \theta$) for AFGL 333 region. The color map is adapted from \cite{2009ApJS..182..143M}, as the polarization map was not re-created in this study.}\label{Measurement of theta major}
\end{figure*}

\subsection{Statistical analysis of the filament polarization data}
In this section, the magnetic field morphology in the filament regions is studied in detail.

\subsubsection{Magnetic field in filament regions:}

In this study, the data were sampled at a resolution of 10''; in certain instances, it was binned to 20'', as detailed in \cite{2009ApJS..182..143M}. The dataset exclusively comprises regions where significant polarization vectors were detected, adhering to the criteria of \(p/\Delta p > 2\), \(\Delta p < 4\%\), and \(I > 0\). To investigate the morphology of magnetic fields within filament regions, we selected six filaments that were observed using the JCMT/SCUPOL at a wavelength of 850 $\mu$m and a resolution of 15'' \citep{2009ApJS..182..143M}.

\subsubsection{Histogram of $\theta_B$ and scatter plot of p vs. $\theta_B$ and KDE plot of the difference between $\theta_{major}$ and $\theta_{B_{mean}}$:}

The histogram illustrating the position angle ($\theta_B$) of the B-vector, along with the scatter plot depicting the degree of polarization (p) against $\theta_B$ (Figure 8), was constructed following the methodology employed for the core regions. These histograms reveal the distribution of $\theta_B$, with the blue dotted line indicating the mean value of the magnetic field position angle, denoted as $\theta_{B_{mean}}$. The scatter plot presents the degree of polarization at varying angles $\theta_B$, with error bars representing the uncertainties associated with the corresponding p and $\theta_B$ values.

The scatter plots, inclusive of error bars, demonstrate that the majority of data points are confined within the following ranges: $113^\circ < \theta_B < 160^\circ$ and $3\% < p < 10\%$ for the AFGL 333 region; $90^\circ < \theta_B < 130^\circ$ and $1\% < p < 16\%$ for the OMC-1 region; $0^\circ < \theta_B < 100^\circ$ and $0.8\% < p < 8\%$ for the OMC-2 \& OMC-3 regions; $125^\circ < \theta_B < 177^\circ$ and $1\% < p < 10\%$ for the G011.11-0.12 region; and $75^\circ < \theta_B < 175^\circ$ and $0.6\% < p < 7\%$ for the DR 21 region. The angle $\theta_{major}$, defined as the angle between the major axis of the filament and the reference north-south direction, represents the longest dimension of the filament, which is delineated within the outermost contour of the respective region, as described in \cite{2009ApJS..182..143M}. Figure 9 illustrates how the $\theta_{major}$ and $\theta_{B_{mean}}$ are measured for the filament AFGL 333.

The mean value of $\Delta\theta$ (defined as $|\theta_{B_{mean}} - \theta_{major}|$), indicating the deviation of the mean B-field direction from the major axis of the filament, is computed to be $75^\circ \pm 7^\circ$ as derived from Table 3. This finding suggests that the major axis of the filament is nearly perpendicular to the direction of the mean magnetic field.

This observation is further corroborated by the KDE plot (Figure 10), which illustrates the peak value of the KDE of $(\Delta\theta)_{mean}$ at $70^\circ$, indicated by the blue dashed line.

The dispersion of $\Delta \theta$ values within the filament regions is approximately $29.83^\circ$. This substantial dispersion highlights the random alignment of the B-field position angles within these filamentous structures.

\vspace{0.2em}

\begin{table*}
\centering
\scriptsize
    \caption{Table to calculate $\Delta\theta$ = $|\theta_{major} - \theta_{B_{mean}}|$ for individual filament region}\label{tableExample} 
    \begin{tabular}{lccccccc}
    \hline
    Name & {RA (J2000)} & {DEC (J2000)} & {Object type} & $\theta_{major}$($^{\circ}$) &  $\theta_{B_{mean}}$($^{\circ}$) & $\Delta\theta$($^{\circ}$) & {Distance (kpc)}\\\hline
    AFGL 333 & {02 28 08.81} & {+61 29 25.0} & {SFR (HM)} & 19 & $111 \pm {7}$ & $92 \pm {7}$ & {1.95} \citep{xu2006distance}\\
    OMC-1 & {05 35 14.5} & {-05 22 33.0} & {SFR (HM)} & 10 & $115 \pm {4}$ & $105 \pm {4}$ & {$0.414 \pm 0.007$} \citep{menten2007distance}\\
    OMC-2 $\&$ OMC-3 & {05 35 26.9} & {-05 09 58} & {SFR (LM)} & 166 & $69 \pm {7}$ & $97 \pm {7}$ & {$0.414 \pm 0.007$} \citep{menten2007distance}\\
    NGC 2068 & {05 46 37.64} & {+00 00 33.1} & {SFR (LM)} & 62 & $77 \pm {8}$ & $15 \pm {8}$ & {0.400} \citep{anthony1982h}\\
    G011.11-0.12 & {18 10 33.99} & {-19 21 36.9} & {SFR (HM)} & 55 & $126 \pm {8}$ & $71 \pm {8}$ & {3.6} \citep{carey2000submillimeter}\\
    DR 21 & {20 39 01.50} & {+42 19 38.0} & {SFR (HM)} & 176 & $108 \pm {9}$ & $68 \pm {9}$ & {3} \citep{campbell1982far}\\
    \hline
    \end{tabular}
    \begin{tablenotes}
        \small
       The values for $\theta_{B_{mean}}$ presented in this table are derived from JCMT/SCUPOL observations, while the classifications for "Object Type", "Distance" and the locations of the regions (RA and DEC)  are adopted from \cite{2009ApJS..182..143M}. Right Ascension (RA) and Declination (DEC) are specified in the conventional units of hours, minutes, and seconds for RA, and degrees, arcminutes, and arcseconds for DEC. The references for the distances associated with the regions taken from \cite{2009ApJS..182..143M} are presented in parentheses adjacent to the corresponding values. The star-forming regions (SFRs) are categorized by mass as high-mass (HM), low-mass (LM), and intermediate-mass (IM), with these designations indicated in parentheses following the respective SFRs.
    \end{tablenotes}
\end{table*}

\begin{figure*}[h]
\centering
\includegraphics[width=1.65\columnwidth]{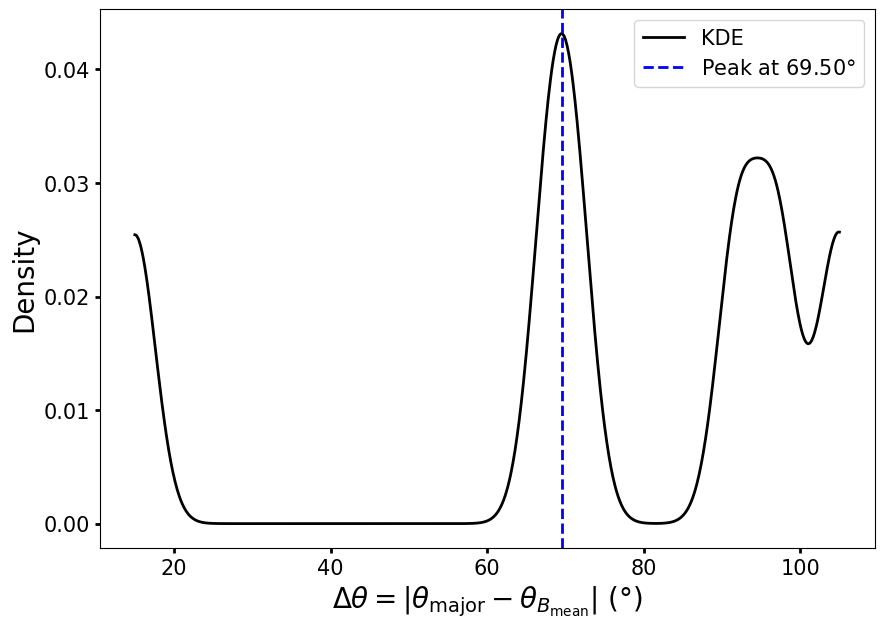}
\caption{KDE plot of the angular difference, $\Delta\theta$ (= $|\theta_{B_\text{mean}} - \theta_\text{major}|$), for the filament regions. The blue dashed line denotes the peak of the KDE curve for $\Delta\theta$, which occurs at 69.50$^\circ$. This suggests that the plane-of-sky component of the mean magnetic field is nearly perpendicular to the major axis of the filamentary structures.}\label{KDE plot for theta major and theta B mean}
\end{figure*}


\subsubsection{Histogram of the difference between $\theta_{major}$ and $\theta_{B_i}$:}

The histogram depicting the difference between $\theta_{major}$ and $\theta_{B_i}$ (Figure 11) elucidates the extent to which the observed position angle of the magnetic field, $\theta_{B_i}$, deviates from the major axis of each filament region.

For all filament regions analyzed, the observed differences between $\theta_{major}$ and $\theta_{B_i}$ indicate a predominant trend where the major axis of the filament is nearly perpendicular to the corresponding observed position angle of the magnetic field ($\theta_{B_i}$).

\vspace{1em}
\begin{figure*}[h]
\centering
\includegraphics[width=2\columnwidth]{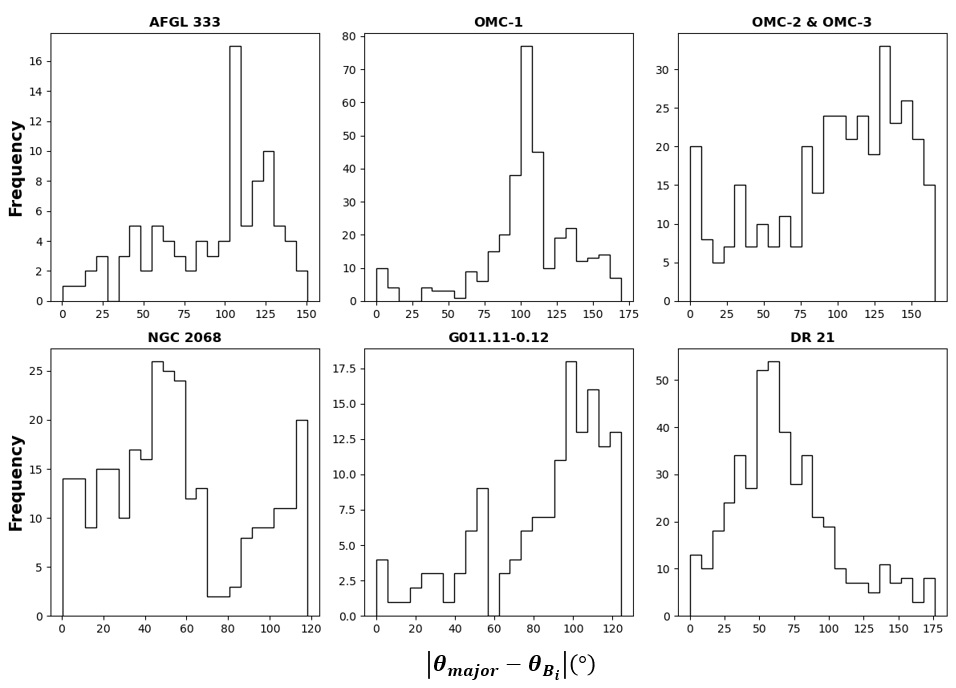}
\caption{Histogram of the difference between $\theta_{major}$ and $\theta_{B_i}$ for the filament regions. These histograms show that for all the filament regions the major axis of the filament is almost perpendicular to the observed position angle of the magnetic field ($\theta_{B_i}$). }\label{Histogram for theta major and theta Bi}
\end{figure*}

\section{Discussion}
\subsection{Star formation in core region}

In the core region, two distinct categories of particles are present: charged particles, such as electrons and ions, and neutral particles, including atoms and molecules. The dynamics of these particles differ fundamentally due to their interactions with the magnetic field. Charged particles are constrained to traverse along the magnetic field lines, following a helical trajectory, while neutral particles, experiencing no magnetic force, possess the freedom to move in any direction.

As depicted in Figure 12, both charged and neutral particles enter the core region by aligning with the magnetic field lines. However, only the neutral particles can approach the core from various directions. This results in a greater degree of contraction along the magnetic field lines, causing the minor axis of the core region to align closely with the direction of the magnetic lines of force \citep{smith2004origin, soam2015magnetic}. This phenomenon, known as "ambipolar diffusion," plays a pivotal role in star formation. The extent of contraction along the direction of the magnetic field is contingent upon the degree of ionization; a lower level of ionization leads to a correspondingly diminished contraction.

Through the process of ambipolar diffusion, a core that is initially sub-critical—characterized by magnetic energy that surpasses gravitational energy, thus preventing collapse—can transition to a super-critical state, where gravitational forces dominate over magnetic influence, allowing for collapse and the potential initiation of star formation. Figure 12 illustrates a cartoon model of ambipolar diffusion, elucidating the alignment of the mean magnetic field direction with the minor axis of the core and demonstrating how the infall of particles within the core region contributes to the onset of star formation.

\begin{figure*}[h]
\centering
\includegraphics[width=2\columnwidth]{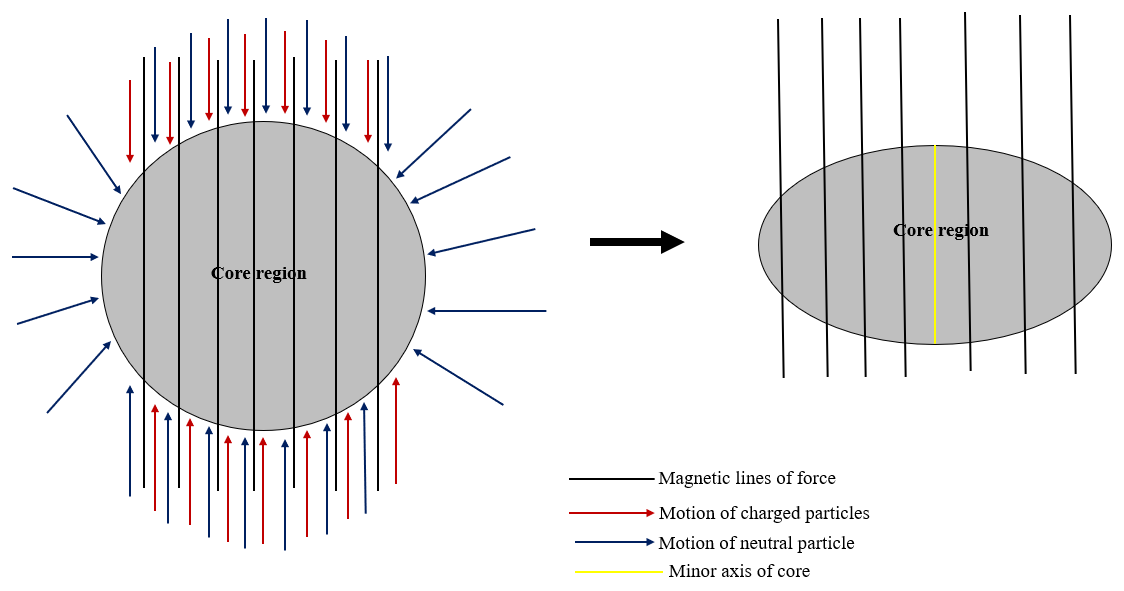}
\caption{Cartoon model of ambipolar diffusion and contraction of the core region along the magnetic field direction.}\label{Cartoon model for ambipolar diffusion}
\end{figure*}

\subsection{Star formation in filament region}

Filamentary structures typically arise in regions characterized by weaker magnetic fields that are influenced by the surrounding gas flow. \cite{gomez2018magnetic} have demonstrated the U-shaped morphology of the magnetic field in proximity to the filament. In these regions, the gas flows align with the direction of the magnetic field, thereby necessitating that the plane of the sky component of the magnetic field remains perpendicular to the long axis of the filament.

In our investigation, we conducted a comprehensive statistical analysis of the magnetic field morphology within the plane of the sky. Our findings indicate that the orientation of the magnetic field is predominantly parallel to the minor axis of the core region while remaining nearly perpendicular to the major axis of the filament.

\section{Conclusion}
The statistical analysis of the combined polarization measurements in the cores and filaments is summarized as follows:
\begin{itemize}
    \item The statistical evaluation of the magnetic field morphology within core regions reveals that the plane of the sky component of the mean magnetic field is predominantly parallel to the minor axis of the core regions.
    \item In several core regions, the outflow direction aligns closely with the large-scale mean magnetic field; however, it exhibits a near-perpendicular relationship with the mean magnetic field measured on smaller scales.
    \item The mean magnetic field direction within core regions demonstrates variability between large and small scales, and this misalignment contributes to a reduction in the degree of polarization (p).
    \item The analysis of magnetic field morphology in filament regions indicates that the plane of the sky component of the mean magnetic field is largely perpendicular to the major axis of the filament regions.
    \item In filament regions, the major axis of the filament is almost perpendicular to the observed B-vector position angle values.
\end{itemize}

\section*{Acknowledgements}
This research has made use of the SIMBAD database,
operated at CDS, Strasbourg, France. We also
acknowledge the use of NASA’s SkyView facility
(http://skyview.gsfc.nasa. gov) located at NASA Goddard
Space Flight Center. We acknowledge the usage of archival data from JCMT/SCUPOL and CARMA/TADPOL surveys.

P.P acknowledges Dr. Archana Soam for her invaluable guidance and insightful suggestions, which greatly enriched this project and contributed significantly to its successful completion.

$Software$: APLpy \cite{robitaille2012aplpy}; \cite{ginsburg2019astroquery}, Astropy  \cite{green2018dustmaps}

\bibliography{biblio}{}

\end{document}